\documentclass[fleqn,usenatbib]{mnras}

\usepackage{newtxtext,newtxmath}

\usepackage[T1]{fontenc}
\usepackage{ae,aecompl}

\usepackage{graphicx}	
\usepackage{amsmath}	
\usepackage{amssymb}	

\usepackage{etoolbox}
\makeatletter
\patchcmd\@combinedblfloats{\box\@outputbox}{\unvbox\@outputbox}{}{%
   \errmessage{\noexpand\@combinedblfloats could not be patched}%
}%
 \makeatother

\AtBeginShipout{%
  \ifnum\value{page}>2 %
    \typeout{* Additional boxing of page `\thepage'}%
    \setbox\AtBeginShipoutBox=\hbox{\copy\AtBeginShipoutBox}%
  \fi
}

\title[$\alpha$ Centauri A as a potential stellar model calibrator]
{$\alpha$ Centauri A as a potential stellar model calibrator: establishing the nature of its core }

\author[B. Nsamba et al.]{
B. Nsamba,$^{1,2}$\thanks{E-mail: benard.nsamba@astro.up.pt}
M. J. P. F. G. Monteiro, $^{1,2}$
T. L.  Campante, $^{1,2}$
M. S. Cunha, $^{1,2}$ 
\newauthor and S. G. Sousa, $^{1}$
\\
$^{1}$Instituto de Astrof\'{\i}sica e Ci\^{e}ncias do Espa\c{c}o, Universidade do Porto,  Rua das Estrelas, PT4150-762 Porto, Portugal\\
$^{2}$Departamento de F\'{\i}sica e Astronomia, Faculdade de Ci\^{e}ncias da Universidade do Porto, PT4169-007 Porto, Portugal
}

\date{Accepted 2018 May 23. Received 2018 May 10; in original form 2018 March 23}

\pubyear{2018}

\begin{document}
\label{firstpage}
\pagerange{\pageref{firstpage}--\pageref{lastpage}}
\maketitle

\begin{abstract}
Understanding the physical process responsible for the transport of energy in the core of $\alpha$ Centauri A is of the utmost importance if this star is to be used in the calibration of stellar model physics. Adoption of different parallax measurements available in the literature results in differences in the interferometric radius constraints used in stellar modelling. Further, this is at the origin of the different dynamical mass measurements reported for this star. With the goal of reproducing the revised dynamical mass derived by \citeauthor{Pour2016}, we modelled the star using two stellar grids varying in the adopted nuclear reaction rates. Asteroseismic and spectroscopic observables were complemented with different interferometric radius constraints during the optimisation procedure. Our findings show that best-fit models reproducing the revised dynamical mass favour the existence of a convective core ($\gtrsim$ 70\% of best-fit models), a result that is robust against changes to the model physics. If this mass is accurate, then $\alpha$ Centauri A may be used to calibrate stellar model parameters in the presence of a convective core.
\end{abstract}
 


\section{Introduction}
Stellar physicists have for decades yearned for a star more massive than the Sun with a range of precisely measured observables, namely, spectroscopic parameters, an interferometric radius, asteroseismic properties, and a dynamical mass measurement. This stems from the fact that stellar model physics (e.g. mixing length parameter, treatment of the initial helium mass fraction, surface element abundances etc.) is often calibrated based on the Sun and used in the modelling of other stars (e.g. \citealt{gaard,Asplund,Bonaca,Vorontsov,Aguirre,Aguirre1}). This is a reasonable approach for stars within the same mass range and with a similar metal content as the Sun. For more massive stars, however, this may not hold, since their internal structure significantly differs from that of the Sun.

$\alpha$ Centauri A presents a unique opportunity to improve our understanding of the underlying physical processes taking place in stars slightly more massive than the Sun. This is due to a number of reasons: (i) $\alpha$ Centauri A is one of the components of the closest binary system to the Sun, having well determined orbital parameters \citep{Pour2016,Kervella2017}\footnote{Hereafter, we note  \citet{Pour2002} as P02, \citet{Kervella2003} as K03, \citet{Kervella2016} as K16, \citet{Pour2016} as P16, \citet{Kervella2017} as K17, and \citet{derhjelm} as S99.}. A dynamical mass measurement is available for both components of the binary (P02; P16; K16). (ii) Precise parallax measurements are available and have been used to yield a distance to the star (S99; K16). This distance has been combined with an interferometric measurement of the star's angular diameter to obtain its radius (K03; P16; K17). (iii) Spectroscopic parameters (e.g. effective temperature, metallicity etc.) are readily available. (iv) Several ground-based campaigns have been conducted in order to obtain asteroseismic data for this star \citep{Bouchy,Bed,Bazot2007,Meulen2010}. The combination of the above set of observables has thus the potential to place tight constraints on the stellar modelling process and help generating best-fit models\footnote{In this work, we refer to a set of models that reproduce a specific set of spectroscopic, seismic, and interferometric constraints as best-fit models.} that can be used in understanding the internal structure of $\alpha$ Centauri A with unprecedented precision.

The dynamical mass of $\alpha$ Centauri A is estimated to span the range [1.10, 1.13] M$_\odot$. Stellar models constructed at solar metallicity within this mass range may display a convective core while on the main sequence, making core overshoot a crucial process to be included in stellar model grids. For this reason, efforts have been made throughout the years to unveil the nature and core properties of $\alpha$ Centauri A using the above set of observables \citep{Monta,Baz}.
 \begin{table*}
\centering 
\caption{Stellar parameters from different literature sources. $\theta_{\rm LD}$ is the angular diameter. The sources for the parallax and angular diameter measurements used in deriving the dynamical mass and interferometric radius are indicated as superscripts. $^a$ denotes S99, $^b$ S99 and K03, $^c$ K03 and K16, $^d$ K16, $^e$ K03 and P16, $^f$ P16, and $^g$ K16 and K17.}
\begin{tabular}{lcccccc}        %
\hline 
Parameter	                & S99		             & P02                        &	K03                     &   P16  						&  K16                                            & K17                  \\
\hline
$\theta_{\rm LD}$ (mas)     & --					&  --				           &  8.511 $\pm$ 0.020      &    --                           &   --                                     & 8.502 $\pm$ 0.038   \\
Parallax (mas)	            & 747.1$\pm$ 1.2   	    &   --                         &  --					  &   743 $\pm$ 1.3             &   747.17 $\pm$ 0.61                          &  --                    \\
Radius (R$_\odot$)  	    &  --                   &   --                        &  1.224 $\pm$ 0.003$^b$    &  1.231 $\pm$ 0.0036$^e$      &  1.2234 $\pm$ 0.0053$^c$                    &  1.2234 $\pm$ 0.0053$^g$    \\
Mass (M$_\odot$)	        &  --                   &  1.105 $\pm$ 0.0070$^a$         & --                    &  1.133 $\pm$ 0.0050$^f$      &  1.1055 $\pm$ 0.0039$^d$                     & --                      	     \\
\hline                                   
\end{tabular}\\
\label{obs}
\end{table*}
A radius measurement (with a precision of about 1\%), when combined with spectroscopic and seismic constraints, has been shown to yield stellar masses with a precision of about 1\%  \citep{Creevey}. $\alpha$ Centauri A has been modelled by several teams, who adopted the interferometric radius of K03 (i.e., 1.224 $\pm$ 0.003 $\rm R_\odot$) as well as complementary spectroscopic and seismic data \citep{Thou,Monta,Baz}. They were able to reproduce the dynamical mass derived by P02. However, differences in the parallax measurements available in the literature inevitably lead to differences in the interferometric radius measurements. This also yields different dynamical mass measurements for the star (see Table~\ref{obs}).

P16 combined radial velocity data from HARPS (High Accuracy Radial velocity Planet Searcher) spanning a period of ten years with data obtained with the Coud\'e Echelle Spectrograph (CES), further complemented by visual observations \citep{Pour1999}, to generate a revised parallax measurement. This revised parallax places the star at a slightly different distance compared to that measured by S99. This led to the revision of the dynamical mass of $\alpha$ Centauri A by P16. The interferometric radius was also revised by combining the new parallax with the angular diameter measurement from K03 (see Table~\ref{obs}). K16 also computed orbital parameters for $\alpha$ Centauri A by combining the same high precision radial velocity data set as P16 with their latest astrometric measurements. They found most of the orbital elements to be commensurate with those found by P16. However, they found a smaller semi-major axis, $a$, emerging from the new astrometry (see table~1 in K16). When $a$ was combined with the high precision radial velocities, they obtained a parallax measurement similar to the one found by S99 but larger than that of P16 (see Table~\ref{obs}). Differences are also evident in the derived dynamical masses and interferometric radius measurements of P16 and K16.

With regard to the nature of the core of $\alpha$ Centauri A, no definitive answer has been reached yet. \citet{Monta} attribute this to the quality of the seismic data available by then (see \citealt{Bed2004}). \citet{Bazot2007} obtained a new set of seismic data using the HARPS spectrograph, having investigated the nature of the core of $\alpha$ Centauri A in \citet{Baz}. They found that approximately 40\% of their best-fit models, which reproduce the dynamical mass derived by P02, possess convective cores. However, the authors point out that this number depends sensitively on the nuclear reaction rates adopted in their models. We note that they used the interferometric radius derived by K03 in their optimisation procedure. 

\citet{Meulen2010} have generated the state-of-the-art seismic data set for $\alpha$ Centauri A by combining the  radial  velocity  time  series  obtained with three  spectrographs  in  Chile  and  Australia (namely, CORALIE, UVES, and UCLES). Here, we adopt this data set and assess the occurrence of best-fit models with convective cores when trying to reproduce the dynamical masses derived by both P16 and K16. The paper is organised as follows.  In Sect.~\ref{Grid}, we describe our stellar models and the parameter ranges used in the construction of the model grids. In Sect.~\ref{observables}, we present the sets of observables used in the optimisation procedure, while the main results are discussed in Sect.~\ref{results}. Section \ref{con} contains our conclusions.
\section{Stellar Model Grids}
\label{Grid}

We constructed two grids (A and B) of stellar models using MESA (Modules for Experiments in Stellar Astrophysics) version 9793 \citep{Pax3}. These grids differ only in the adopted nuclear reaction rates (see Table \ref{rates} for details). \citet{Baz} found the occurrence of models of $\alpha$ Centauri A with convective cores to vary mainly due to the choice of nuclear reaction rates, in particular that of the $^{14}{\rm N}(p,\gamma)^{15}{\rm O}$ reaction. This reaction rate is crucial for the CNO (carbon-nitrogen-oxygen) cycle and its variation is expected to significantly affect the chances of a model developing a convective core. 
We therefore varied the nuclear reaction rates in order to test the robustness of the occurrence of best-fit models with convective cores when using different observational constraints (see Sect.~\ref{observables}).  
\begin{table}
\centering 
\caption{Main features of the stellar model grids adopted in this work.}
\begin{tabular}{lccr}        %
\hline 
Grid	             & Reaction Rates		&  Core Overshoot	&	Diffusion	       \\
\hline
A	             & JINA REACLIB	   		&  Yes				&   Yes							\\
B  	  		     & NACRE   	       		&  Yes  			&   Yes							\\
\hline                                   
\end{tabular}
\label{rates}
\end{table}
Grid A employs nuclear reaction rates from JINA REACLIB 
(Joint Institute for Nuclear Astrophysics Reaction Library) version 2.2 \citep{Cyburt}. It should be noted that grid A uses specific rates for 
$^{14}{\rm N}(p,\gamma)^{15}{\rm O}$ and $^{12}{\rm C}(\alpha, \gamma)^{16}{\rm O}$ described by \cite{Imbriani} and \cite{Kunz}, respectively.
Grid B employs nuclear reaction rates as obtained from tables provided by the NACRE (Nuclear Astrophysics Compilation of Reaction Rates) collaboration \citep{Angulo}. Furthermore, element diffusion is a relevant transport process in low-mass stars, i.e., below $\sim$1.2 $\rm M_\odot$ (e.g. \citealt{2018Nsamba}), and was therefore included in our model grids. Core overshoot becomes a vital process once a stellar model develops a convective core and was included in such models.

The version of MESA used in this paper adopts the 2005 update of the OPAL equation of state \citep{Rogers}. Opacities from OPAL tables \citep{Iglesias} were used at high temperatures while tables from \cite{Ferguson} were adopted at lower temperatures. 
We used the surface chemical abundances of \citet{Grevesse} with a solar metal mass fraction value of 0.0169.
The standard Grey--Eddington atmosphere  was used to describe the surface boundary (it integrates the atmosphere structure from the photosphere down to an optical depth of $10^{-4}$). Convection was described using the mixing length theory \citep[MLT;][]{Vitense} while element diffusion was  implemented according to \citet{Thoul}. Element diffusion includes gravitational settling and chemical diffusion.
The helium-to-heavy metal enrichment relation was used to determine the helium mass fraction ($\rm Y$). The ratio $\Delta \rm Y  / \Delta \rm Z$ = $\rm 2$ \citep{Chiosi} was used, while $\rm Z_{0}$ = 0.0 and $\rm Y_{0}$ = 0.2484 were set based on the big bang nucleosynthesis \citep{Cyburt2003}.

Evolutionary tracks are varied in mass, $\rm M$, metal mass fraction, $\rm Z$, mixing length parameter, $\alpha_{\rm mlt}$, and 
core overshoot parameter, $\rm f$. We used the exponential diffusive overshoot recipe in MESA when describing core overshoot mixing. The diffusion coefficient (D$_{\rm c}$) in the overshoot region is expressed as \citep{Herwig}:
\begin{equation}
\rm D_{\rm c} = \rm A_{\rm 0} exp\left(   \frac{-2z}{\rm f \cdot H_{\rm p}}    \right) ~~,
\label{rad}
\end{equation}
where A$_{\rm 0}$ is the diffusion coefficient in the convectively unstable region near the convective boundary determined using MLT, H$_{\rm p}$ is the pressure scale height, and $\rm z$ is the distance from the edge of the convective zone. 

Grid parameter ranges are: $\rm M \in$ [1.0, 1.2] M$_\odot$ in steps of 0.01 M$_\odot$, $\rm Z \in$ [0.023, 0.039] in steps of 0.001, $\alpha_{\rm mlt} \in$ [1.3, 2.5] in steps of 0.1, and $\rm f \in$ [0, 0.03] in steps of 0.005.
We kept models starting from the ZAMS (zero age main-sequence; defined as the point along the evolutionary track where the nuclear luminosity is 99\% of the total luminosity) to the end of the sub-giant evolution stage. 
Using GYRE \citep{Townsend}, we generated adiabatic oscillation frequencies for spherical degrees $l$ = 0, 1, 2, and 3 for each model.
\section{Observational Constraints and Optimisation Procedure}
\label{observables}

We downloaded a few high S/N, individually reduced HARPS observations of $\alpha$ Centauri A, which were combined to generate a final spectrum for subsequent analysis. Spectroscopic parameters (i.e., effective temperature, $T_{\rm eff}$, and metallicity, $\rm [Fe/H]$) were derived based on the analysis of the equivalent widths of Fe I and Fe II lines measured with ARES \citep{Sousa2007, Sousa2015} and assuming LTE (Local Thermodynamical Equilibrium). We used the MOOG code \citep{Sneden} and a set of plane-parallel ATLAS9 model atmospheres \citep{Kurucz1993} in our analysis, as described in \citet{Sousa2011}. For more details on the combined ARES+MOOG method, we refer the reader to \citet{Sousa2014}. We obtained $T_{\rm eff} = 5832 \pm 62$ $\rm K$ and $\rm [Fe/H] =  0.23 \pm 0.05$ dex. 

Using the angular diameter measurement of K17 together with the parallax measurement of P16, we revised the interferometric radius of $\alpha$ Centauri A by means of the expression \citep{Ligi}:
\begin{equation}
 R (R_\odot) = \frac{\theta_{\rm LD} \times \rm d{[\rm pc]}}{9.305} ~~,
\label{inter_R}
\end{equation}
where $\rm d{[\rm pc]}$ is the distance to the star expressed in parsec. 
 \begin{table}
\centering 
\caption{Spectroscopic and interferometric constraints adopted during the optimisation procedure.}
\begin{tabular}{lccr}        %
\hline 
Run	             & $T_{\rm eff}$ (K)		&  [Fe/H] (dex)		&	Radius	       \\
\hline
1	             & 5832 $\pm$ 62   	   		&  0.23 $\pm$ 0.05	&  1.231 $\pm$ 0.0036	\\
2  	  		     & 5795 $\pm$ 19   	        &  0.23 $\pm$ 0.05  &  1.2234 $\pm$ 0.0053	\\
\hline                                   
\end{tabular}
\label{adopted}
\end{table}
We find the revised interferometric radius to be 1.230 $\pm$ 0.0056 $\rm R_\odot$. This is in agreement with the value obtained by P16 within 1$\sigma$.

We will be considering two optimisation runs in this work (Run 1 and Run 2) depending on the set of observables adopted (see Table~\ref{adopted}). The value of $T_{\rm eff}$ used in Run 1 was derived in this work while the interferometric radius is from P16. For self-consistency, Run 2 uses the interferometric radius and $T_{\rm eff}$ from K16. The value of [Fe/H] used in both runs is the one derived here.

The set of observables in Table~\ref{adopted} was complemented with seismic data (i.e., individual oscillation frequencies) from \citet{Meulen2010}. 
We treated modes exhibiting rotational splittings in the same way as \citet{Meulen2010}, i.e., by taking their average and summing the associated uncertainties in quadrature. This is based on the assumption that such splittings are symmetric. The combined-term, surface frequency correction method of \cite{Ball} was used to handle the offset between observed and model frequencies \citep{Dziembowski}. This method has been shown to yield the least internal systematics in stellar mass, radius, and age when compared to other methods (for details, see \citealt{2018Nsamba}).

Finally, we used AIMS\footnote{http://bison.ph.bham.ac.uk/spaceinn/aims/} (Asteroseismic Inference on a Massive Scale; \citealt{Reese}) to generate a representative set of models reproducing the set of asteroseismic, spectroscopic, and interferometric constraints (as per above). The mean and standard deviation of the posterior probability distribution functions (PDFs) are taken as estimates of the modelled stellar parameters and their uncertainties, respectively.
\section{Results}
\label{results}

\begin{figure*}
\minipage{0.33\textwidth}
  \includegraphics[width=\linewidth]{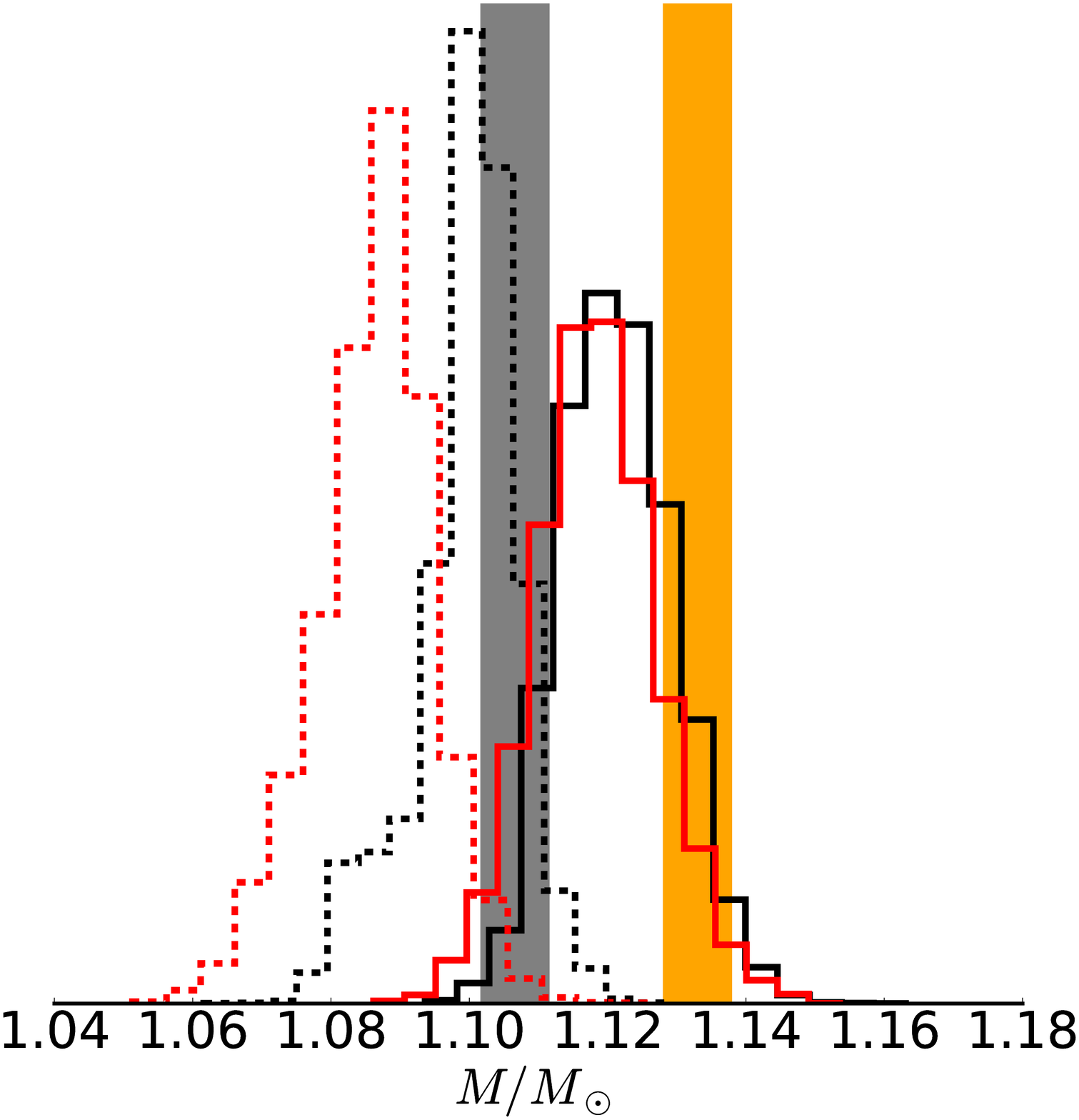}
\endminipage\hfill
\minipage{0.33\textwidth}
  \includegraphics[width=\linewidth]{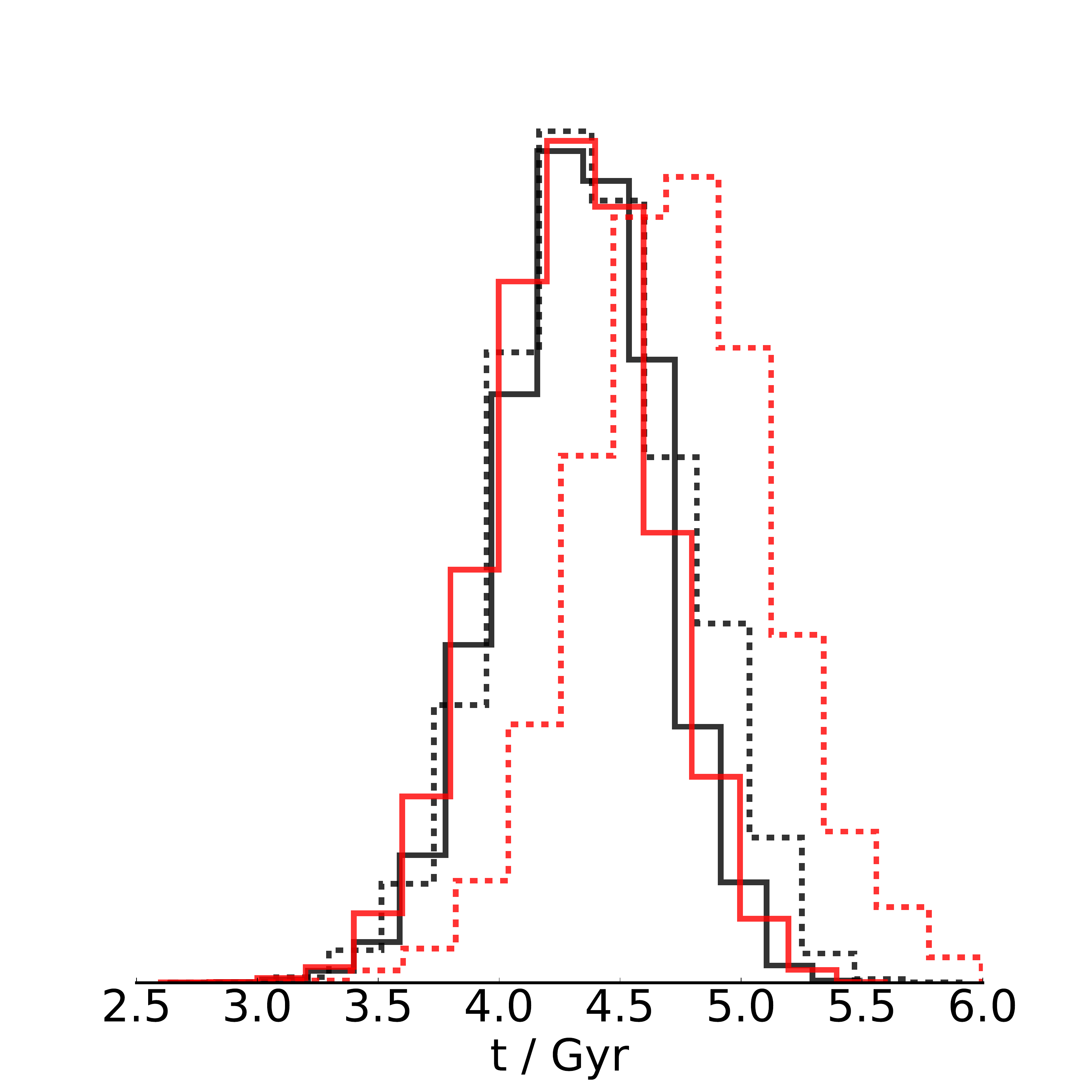}
\endminipage\hfill
\minipage{0.33\textwidth}%
  \includegraphics[width=\linewidth]{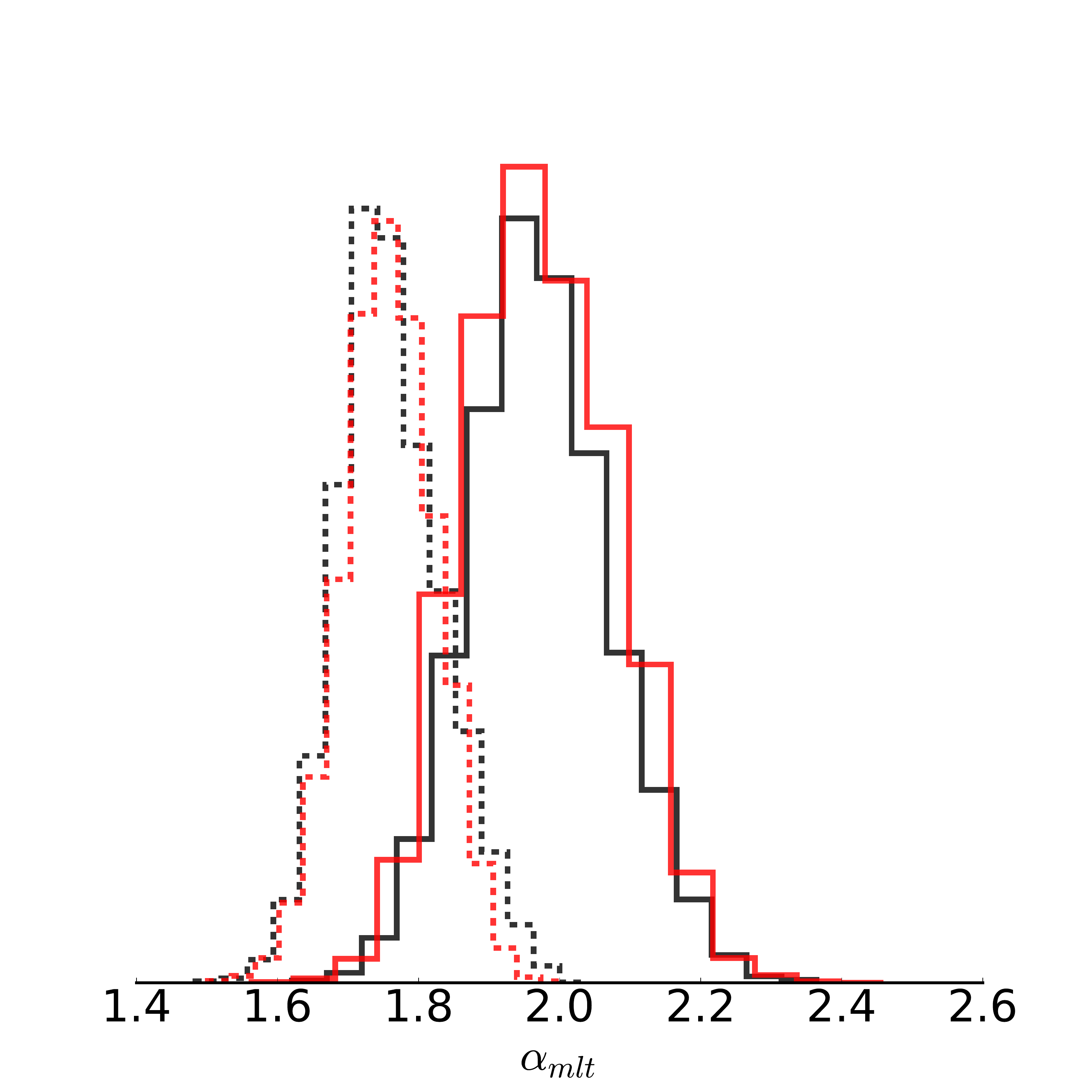}
\endminipage
\caption{Posterior PDFs of stellar parameters obtained using grid A (red) and grid B (black). Solid lines show results from Run 1 while dashed lines are for Run 2. Grey and orange bars correspond to the dynamical mass measurements (and associated uncertainties) of K16 and P16, respectively.}
 \label{parameters}
\end{figure*}

Results obtained by combining both grids (A and B) and sets of observables (Run 1 and Run 2) are shown in Table \ref{values} and Fig.~\ref{parameters}. Only posterior PDFs showing significant differences are shown in Fig.~\ref{parameters}. 

Results based on the set of observables in Run 2 are consistent within 1$\sigma$. The derived stellar mass is in agreement with the dynamical mass obtained by P02 and K16 (even if at the 2$\sigma$ level when considering grid A). These results are also consistent with those obtained by other modelling teams \citep{Monta,Bazot2012,Baz}. This is because these teams complemented seismic and spectroscopic constraints with the interferometric radius of K03, whose value is in close agreement (within 1$\sigma$) with that used in Run 2.
\begin{table*}
\centering 
\caption{Stellar parameters obtained by combining both grids (A and B) and sets of observables (Run 1 and Run 2).}
\begin{tabular}{lcccccccc}        
\hline 
Grid	& Run 	&	$\rm M$ $(\rm M_\odot)$		&  $\rm t$ (Gyr)       	& $\rm Z_0$   			 & $\rm X_0$   			  & $\alpha_{\rm mlt}$   & $\rm Y_{\rm surf}$ & Convective Core (\%)\\
\hline
 A 		&1 		& 	1.12 $\pm$ 0.01		&  4.30 $\pm$ 0.35 	&  0.034 $\pm$ 0.002  &  0.649 $\pm$ 0.006  & 1.97 $\pm$ 0.10   & 0.282 $\pm$ 0.004  & 70 	\\
        &2		&	1.09 $\pm$ 0.01		&  4.74 $\pm$ 0.40 	&  0.035 $\pm$ 0.002  &  0.648 $\pm$ 0.006  &  1.76 $\pm$ 0.07   & 0.280 $\pm$ 0.004 & 46	\\
\hline 
 B 		&1 		&	1.12 $\pm$	0.01	&   4.32 $\pm$ 0.33	&  0.034 $\pm$ 0.002 &  0.650 $\pm$ 0.007 &	1.97 $\pm$ 0.10  & 0.282 $\pm$ 0.005      & 77	\\
        &2		&	1.10 $\pm$ 0.01		&	4.39 $\pm$ 0.38 &  0.034 $\pm$ 0.002 &  0.650 $\pm$ 0.006 & 1.76 $\pm$ 0.07  & 0.281 $\pm$ 0.003      & 77	\\
\hline 
\end{tabular}
\label{values}
\end{table*}

Results based on the set of observables in Run 1 are also consistent within 1$\sigma$.
The derived stellar mass is now in agreement with the revised dynamical mass of P16. We found similar results when replacing the interferometric radius of P16 with that derived in this work (cf.~Sect.~\ref{observables}). This was expected as both values agree within 1$\sigma$.

When adopting the set of observables in Run 1, grids A and B return similar yields of 70\% and 77\% of best-fit models with convective cores, respectively. A contrasting picture emerges when considering the set of observables in Run 2: 46\% (grid A) versus 77\% (grid B). This is mainly due to the different nuclear reaction rates used in both grids. The reaction rate for $^{14}{\rm N}(p,\gamma)^{15}{\rm O}$ from \cite{Imbriani} used in grid A is lower compared to that from NACRE \citep{Angulo} in grid B. This reduces the chances of having convection as a means of energy transport in the core of stellar models in grid A.
In addition, Run 1 yields more models in the high-mass regime (see Fig.~\ref{parameters}), which increases the chances of the CNO cycle being the main energy production chain, resulting in more models with convective cores. 
We also find models with convective cores to have on average a higher metallicity compared to those with radiative cores. This is consistent with the findings of \citet{Baz}.

In the leftmost panel of Fig.~\ref{parameters}, a shift (although still retaining a 1$\sigma$ agreement) in the posterior PDFs for the stellar mass can be seen (dashed lines or Run 2). This is again due to the change in the nuclear reaction rates. Results based on grid B yield a large fraction of models with convective cores and therefore higher masses compared to results based on grid A, for which a relatively large fraction of models with radiative cores are obtained, thus resulting in lower masses on average. The lower masses obtained in the latter case lead to a slightly higher age (middle panel of Fig.~\ref{parameters}). The mass range of best-fit models obtained in Run 1 is shifted toward higher masses than that of Run 2. Also, the effect of changing the nuclear reaction rates turns out to be less effective in this higher-mass range since most models have developed convective cores. This explains the consistency in the results obtained with both grids when using Run 1.

A noticeable difference in the mixing length parameter, $\alpha_{\rm mlt}$, can be seen between the results based on Run 1 and Run 2 (rightmost panel of Fig~\ref{parameters}). The most probable cause for this, is the different interferometric radius constraint used.  

The percentage of best-fit models with convective cores that reproduce the revised dynamical mass derived by P16 (Run 1) is similar for both grids. However, when reproducing the dynamical mass derived by P02 and K16 (Run 2), the percentage of best-fit models with convective cores varies depending on which grid is used, indicating a strong sensitivity to the nuclear reaction rates adopted \citep[cf.][]{Baz}.
\begin{figure}
  \includegraphics[width=\columnwidth, height= 6cm]{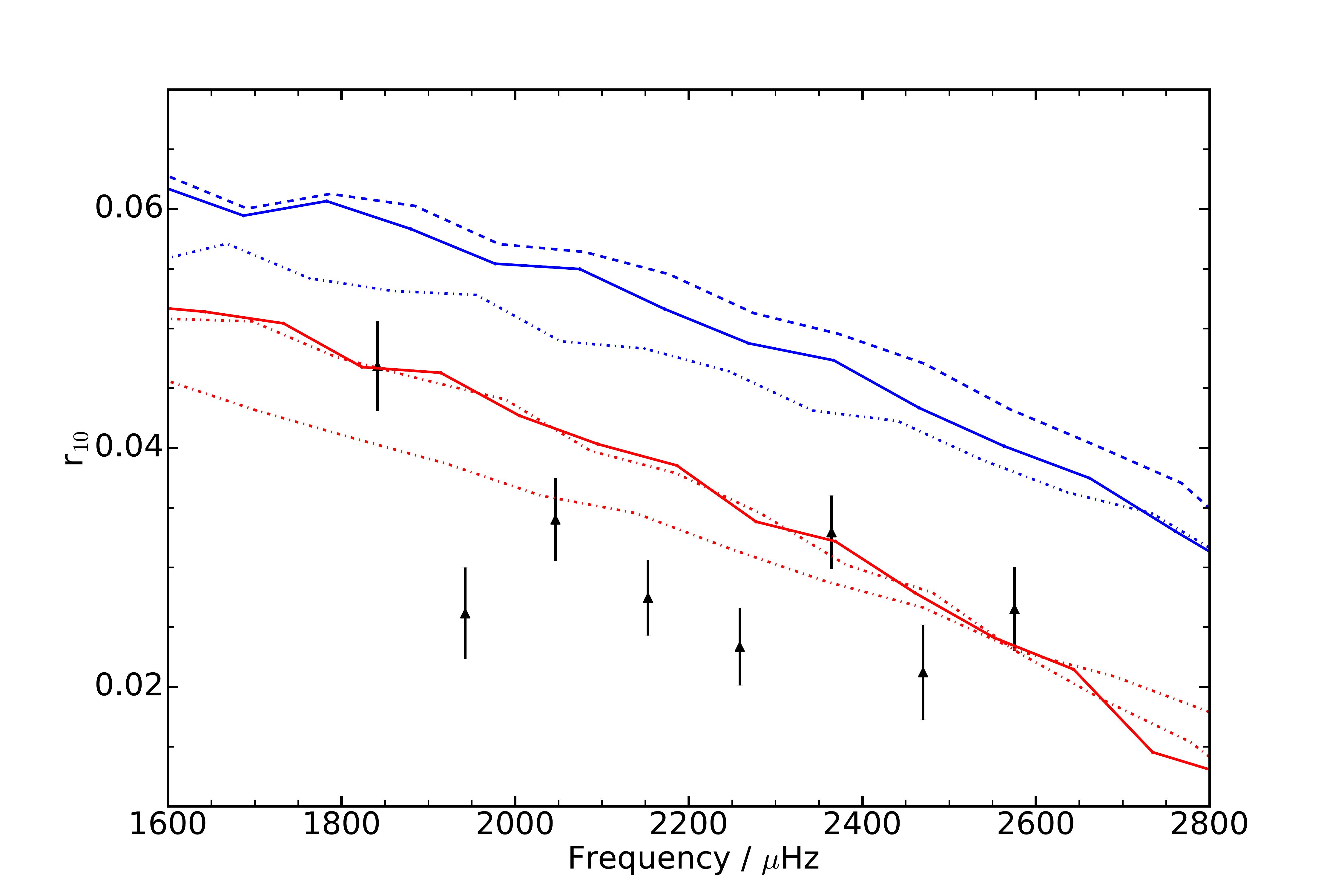}
\caption{Comparison of observed frequency ratios $r_{10}$ (black triangles) with those from models with radiative (blue lines) and convective cores (red lines).}
\label{ratios}
\end{figure}

We further compared the observed frequency ratios, $r_{10}$ (see \citealt{Rox}), to those computed for a handful of representative best-fit models (for Run 1, grid A) having either convective or radiative cores (see Fig.~\ref{ratios}). Frequency ratios are less prone to the outer layers of the star and are therefore reliable indicators of the deep stellar interior conditions. Our findings seem to indicate that models having a convective core lead to a better agreement with the observed $r_{10}$, contrary to what was found by \citet{Meulen2010}. This is not surprising, as we complemented our seismic data with the interferometric radius of P16, which yields models that reproduce well the revised dynamical mass of P16. \citet{Meulen2010}, on the other hand, used models from \cite{Monta}, which reproduce the dynamical mass of P02.

\section{Conclusions}
\label{con}
In this study, we have successfully reproduced the revised dynamical mass of $\alpha$ Centauri A derived by P16 using a forward stellar modelling approach. Our findings show that best-fit models favour the presence of a convective core in $\alpha$ Centauri A, regardless of the nuclear reaction rates adopted in the modelling. We therefore conclude that, if the revised dynamical mass of P16 is accurate, then $\alpha$ Centauri A may be used to calibrate stellar model parameters in the presence of a convective core. Furthermore, the percentage of best-fit models having convective cores that reproduce the smaller dynamical mass published by P02 and K16 varies depending on the choice of nuclear reaction rates.

Our findings further stress the importance of a precise interferometric radius (with a precision better than 1\%) in complementing seismic data with the aim of tightly constraining 
stellar models when adopting a forward modelling approach \citep[cf.][]{Monta,Creevey}.

Seismic diagnostics of the nature of stellar cores based on frequency combinations demand a relative uncertainty on the observed individual frequencies of about $10^{-4}$ \citep[e.g.,][]{Cunha,Brand2014}, commensurate with that obtained from multi-year, space-based photometry \citep{Aguirre2013,Lund2017}. Our results reveal that, for $\alpha$ Centauri A, a median relative uncertainty on the observed individual frequencies of $2.5 \times 10^{-4}$ is sufficient to allow the use of frequency ratios in drawing a distinction -- even if merely qualitative -- between best-fit models with different core properties.

\section*{Acknowledgements}
\footnotesize
This work was supported by Funda\c{c}\~{a}o para a Ci\^{e}ncia e a Tecnologia (FCT, Portugal) through national funds (UID/FIS/04434/2013), by FEDER through COMPETE2020 (POCI-01-0145-FEDER-007672), (POCI-01-0145-FEDER-030389)
and FCT/CNRS project PICS.
BN is supported by FCT through Grant PD/BD/113744/2015 from PhD::SPACE, an FCT PhD programme. MSC is supported by FCT through an Investigador contract with reference IF/00894/2012 and POPH/FSE (EC) by FEDER funding through the program COMPETE. SGS acknowledges support from FCT through Investigador FCT contract No.~IF/00028/2014/CP1215/CT0002 and from FEDER through COMPETE2020 (grants UID/FIS/04434/2013 \& PTDC/FIS-AST/7073/2014 \& POCI-01-0145 FEDER-016880). Based on data obtained from the ESO Science Archive Facility under request number SAF Alpha Cen A 86436.
The authors also acknowledge the anonymous referee  for  the  helpful  and  constructive remarks.


\bibliographystyle{mnras}
\bibliography{mybib} 






\bsp	
\label{lastpage}
\end{document}